\documentclass[pre,aps,showpacs]{revtex4}

\usepackage{amsmath}
\usepackage{amssymb}
\usepackage{graphicx}
\newcommand{\nt}{\tilde{\nu}}
\newcommand{\kt}{\tilde{k}}
\newcommand{\vt}{v_{th}}

\begin{document}

\title{Fluctuating hydrodynamics for dilute granular gases: a Monte Carlo study}

\author{Giulio Costantini}
\affiliation{CNR-ISC and Dipartimento di Fisica, Universit\`a Sapienza - p.le A. Moro 2, 00185, Roma, Italy}

\author{Andrea Puglisi}
\affiliation{CNR-ISC and Dipartimento di Fisica, Universit\`a Sapienza - p.le A. Moro 2, 00185, Roma, Italy}

\begin{abstract}
  We investigate hydrodynamic noise in a dilute granular gas during the
  homogeneous cooling state, by means of a proper application of the
  Direct Simulation Monte Carlo (DSMC) algorithm. The DSMC 
  includes a source of randomization which is not present in Molecular
  Dynamics (MD) for inelastic hard disks. Notwithstanding this
  difference, a fair quantitative agreement is found, including a
  violation of the fluctuation-dissipation relation for the noise
  amplitude of the same order observed in MD. This study suggests that
  deterministic collision dynamics is not an essential ingredient to
  reproduce, up to a good degree of approximation, hydrodynamic
  fluctuations in dilute granular gases.
\end{abstract}


\maketitle

\section{Introduction}

Granular gases represent an important benchmark for theories and
methods of non-equilibrium statistical mechanics~\cite{PL01}. A
granular gas is the idealized counterpart of fluidized granular media,
i.e. collections of $N \sim 10^2 \div 10^3$ macroscopic grains
(typically spheres with diameters $\sigma$ in the range
$10^{-4}\div 10^{-3}m$), contained in volumes $V$ such that the packing
fraction, or occupied volume, is smaller (often much smaller) than
$10\%$: note that the small packing fraction $\sim N\sigma^d/V$ is
still compatible with a non-negligible collision rate, $\sim
N\sigma^{d-1}/V$, where $d$ is the dimensionality of the
system. Within these conditions, the main difference with respect to
an ideal gas is given by inelastic collisions: kinetic energy is lost
during an (hard core) interaction, ruling out many basic assumptions
of equilibrium statistical mechanics and leading to several phenomena
such as: breakdown of energy equipartition~\cite{FM02,MP02b,PMP02},
spontaneous symmetry breaking (vortices and
clustering)~\cite{NEBO97,G99}, ``Maxwell demon''~\cite{Eg99},
``ratcheting'' effects~\cite{CPB07}, and lot more. In both theoretical
investigation and in real world applications, granular ``particles''
are often found in a small number, thousands is a common order of
magnitude, as mentioned before. This makes fluctuations a fundamental
ingredient of granular statistical descriptions~\cite{G99}. While
continuum theories, e.g. granular hydrodynamics and kinetic theory,
describing local averages for density, velocity and energy and their
evolution, have achieved an advanced degree of maturity~\cite{BDKS98},
a good description of noise accompanying such averages is still an
open problem. Near equilibrium, Einstein fluctuation theory, resumed
by Landau and Lifshitz in their fluctuating
hydrodynamics~\cite{LandauFisStat}, based on equilibrium
fluctuation-dissipation relations (FDR)~\cite{KTH91,BPRV08},
simplifies the description of noise for transport processes. In
granular gases, fluctuations do not satisfy, in general,
FDR. The simultaneous presence of external random drivings and diluteness, 
usually guarantees the validity of FDR~\cite{PBL02,BLP04}. FDR is, instead,
violated at packing fractions $>10\%$~\cite{PBV07,VPV08} and in a
non-driven granular gas, even dilute, in the so-called
Homogeneous Cooling State (HCS)~\cite{DB02,BGM08}.

A recent study on dilute inelastic hard particles in the HCS,
with Molecular Dynamics (MD) simulations~\cite{BGM08} and
then an analytical treatment~\cite{BMG09}, have shown that fluctuations
of the hydrodynamic transverse velocity field deviate from the
classical Landau-Lifshitz theory for two reasons: 1) noise is not
white, 2) it does not satisfy the FDR of the 2nd kind. The analytical
study is based on projection operator formalism, starting from the
``pseudo-Liouville'' operator for the deterministic evolution which
only includes free flights and instantaneous hard-core inelastic
collisions at contact. The only statistical averaging is made with
respect to an ensemble of initial conditions, while several hypothesis
are introduced: among the others, spatial homogeneity and a special
time-scaling (inherent to HCS), for one-particle and two-particle
distributions, such that all time-dependence is contained in the
``granular temperature'' $T_g=m\langle {\bf v}^2\rangle/d$. The last
part of this analysis makes use of the Molecular Chaos assumption, which is
reasonable for dilute systems. 
Anyway, one wonders which
ingredients of this sophisticated analysis are essential and which can
be neglected. For instance, the use of Molecular Dynamics (MD) for simulations and
of the rigorous manipulation of the pseudo-Liouville operator as 
starting point of the analytical treatment let us think that the {\em
  deterministic} nature of the hard spheres dynamics is crucial.

Here we present a different model, which in principle could produce
different results, but in the end reveals to be in fair agreement with
the previously cited study. It is a {\em stochastic} model, where
collisions are treated randomly, in sharp contrast with deterministic
dynamics employed in Molecular Dynamics simulations and described by
the pseudo-Liouville operator. It consists of the classical ``Direct
Simulation Monte Carlo'' (DSMC) algorithm~\cite{B94} in its
non-homogeneous form, i.e. placed on a spatial grid to measure modes
of the transverse velocity field, with the use of  ``fictive particles'' 
to guarantee Molecular Chaos (the
algorithm is explained in the Appendix). The inelastic collision rule,
relating post-collisional velocities to pre-collisional ones, and the time-rescaling are the only
ingredients which are conserved with respect to the MD model. Our
question is: are those ingredients sufficient to reproduce the
hydrodynamic noise properties of the deterministic model?

  Similar studies have been performed for hydrodynamic
    fluctuations in the elastic case ~\cite{GMLMC87,MGLC87}.  More
    recently, the DSMC approach has also been applied to fluctuations
    of the global energy fluctuations (a homogeneous, not spatially
    dependent quantity) in inelastic systems~\cite{VPBWT06,BRM07},
    obtaining a good agreement with the amplitude of fluctuations
    measured in MD simulations and explained in terms of projection
    operators applied to the deterministic dynamics~\cite{B04}. A
  similar conclusion was available for that study: stochastic ``Monte
  Carlo'' treatment of collisions is sufficient to reproduce MD noise
  properties. A successive study~\cite{CPB07b} also showed how certain
  quantities (cumulants of fluctuations in the stationary state) may
  be obtained without resorting to projection techniques.

In the present study, in a sense, the observation is even more
striking: the exchange of particles among different ``copies'', needed
to reduce the incidence of re-collisions and finite size correlation
effects, is a strong source of randomization. Surprisingly it does not
affect the hydrodynamic fluctuations.

The organization of the paper is the following: in
Section~\ref{sec:theory} we first describe the model, then in
Section~\ref{sec:numres} we present the results, first verifying
Molecular Chaos and then measuring the hydrodynamic noise in
time-autocorrelation and amplitude. Finally we draw Conclusions. We
have dedicated an Appendix to a detailed description of the DSMC
algorithm used for this work.

\section{The theory} \label{sec:theory}

\subsection{Definition of the model}

This Direct Simulation Monte-Carlo study approximates the dynamics of a dilute system of $N$ smooth
inelastic hard disks of mass $m=1$ and diameter $\sigma$. The inelastic
collisions between the disks $i$ and $j$  change  the
particles velocities following the relations
\begin{eqnarray}
\label{rules}
{\mathbf v}'_i &=&  {\mathbf v}_i-\frac{1+\alpha}{2}(\hat{\sigma}\cdot {\mathbf v}_{ij}) \hat{\sigma} \nonumber \\
{\mathbf v}'_j&=& {\mathbf v}_j+\frac{1+\alpha}{2}(\hat{\sigma}\cdot {\mathbf v}_{ij})\hat{\sigma}
\end{eqnarray}
where ${\mathbf v}'$ corresponds to post-collisional velocity,
$\alpha$ is the coefficient of restitution and ${\mathbf
v}_{ij}={\mathbf v}_i-{\mathbf v}_j$ is the relative velocity. The
term $\hat{\sigma}$ is a random unit vector (see Appendix).  The size
of the system is $L\times L$ and the boundary conditions are periodic.

The main hypothesis used here is that the system, starting in a
uniform equilibrium state with granular temperature $T_g(0)=T_0$,
evolves to the Homogeneous Cooling State (HCS), which is characterized
by a single time-scale measured by the temperature $T_g(t)$: any other
quantity depends on time only through $T_g(t)$. Apparently, as
observed in many previous studies~\cite{BRM04}, in a homogeneous
setting, Molecular Chaos is sufficient to guarantee this hypothesis.

The HCS is unstable against spatial fluctuations: this instability
appears at scales larger than a critical length $L_c(\alpha)$,
therefore it can be avoided by taking the linear size of the system
$L<L_c$~\cite{GZ93}. It is possible to analyze the effects of spatial
fluctuations by deriving mesoscopic equations through a linearization
around the HCS~\cite{NEBO97}. The resulting equations are generally
coupled, but in the Fourier representation the transverse velocity
field results decoupled from the other modes. We are interested, in
particular, in the fluctuations of its largest mode, i.e. of the
smallest wave-number $|{\mathbf k_m}|=2\pi/L$.  We choose the wave vector ${\mathbf k_m}$
parallel to the axis $\hat{x}$ and the quantity above results to be
\begin{equation}
U_{\perp}(t) =\sum_j^N v_{y,j}(t)\exp\Big(i\frac{2\pi x_j(t)}{L}\Big)
\end{equation}
where $v_{y,j}(t)$ is the component of the velocity of the particle
$j$ in the direction $\hat{y}$ and $x_j(t)$ is its coordinate along
the $\hat{x}$ axis.  We want to verify the hypothesis that the
fluctuations of $ U_{\perp}(t)$ obey a linear Langevin
equation,~\cite{NEBO97,BGM08} 
\begin{equation}
\partial_t U_{\perp}(t)=-\nu(t) k_m^2 U_{\perp}(t)+\sqrt{v_{th}(t)N c~\nu(t) k_m^2}\xi(t) 
\label{Langevin}
\end{equation}
where $\nu(t)$ is the kinematic viscosity which, in the HCS, is
proportional to $\sqrt{T_g}$ (see~\cite{BDKS98} for definitions),
and $v_{th}(t)\equiv \sqrt{2T_g/m}$.

The last term in the Eq. (\ref{Langevin}) describes an internal
noise corresponding to the rapid (microscopic) degrees of freedom of the
system. This complex noise is assumed here to be white, Gaussian and with
correlations given by
\begin{eqnarray}
\langle \xi(t)\xi^*(t') \rangle &=&\delta(t-t') \\
\langle \xi(t)\xi(t') \rangle &=&0
\label{noise}
\end{eqnarray}
with $\xi^*(t)$ the complex conjugate of $\xi(t)$. Deviations from
the white noise assumption are expected~\cite{BMG09}, but they
are in general quite small. The most evident consequence of inelasticity
is, instead, the variation of  the dimensionless coefficient $c$ in
Eq. (\ref{Langevin}), which is equal to $1$ at equilibrium
($\alpha=1$); its departure from $c=1$ represents the fact that the
fluctuations of $U_{\perp}(t)$ do not satisfy the FDR.  

Since the system is cooling, the typical velocity of the particles
becomes smaller and smaller, leading to increasing rounding errors and
other numerical problems. These problems can be avoided with the use
of a procedure which, in the HCS, maps the dynamics to a steady state
by means of a time-rescaling~\cite{BRM04,Lu01}. Such procedure is explained
in the next subsection.

\subsection{The stationary representation}

In the HCS, the granular temperature obeys the following equation:
\begin{equation}
 \partial_t T_g(t)+\zeta_H(t)~ T_g(t)=0
\label{temperature}
\end{equation}
where $\zeta_H$ is the cooling rate that results proportional to
$\sqrt{T_g(t)}$. \\
The stationary representation of the HCS~\cite{BRM04} consists in
introducing a new time scale $\tau$ defined by
\begin{equation}
\omega_0\tau=\ln\frac{t}{t_0}\label{resc}
\end{equation}
with $\omega_0$ ant $t_0$ arbitrary constants, implying the definition
of rescaled velocities $\tilde{\mathbf v}(\tau)={\mathbf v}(t)\omega_0
t $. It is easy to see~\cite{BRM04} that observing the system on this
new time scale is equivalent to apply a {\em positive} continuous drag
to all particles $\partial_\tau \tilde{\mathbf v}(\tau)=\omega_0  \tilde{\mathbf v}(\tau)$.  This naturally leads to define also
the rescaled analogous of $U_\perp(t)$: $W_{\perp}(\tau)=U_\perp(t)\omega_0
t$.\\
We tune $\omega_0$ in order to have $T_g(\tau \to \infty)=T_0$ (so
that the length of transients is reduced), a result which is obtained
by taking $\omega_0=\zeta_H(0)/2$. In the steady state, we simplify the
notation using $v_{th}=v_{th}(0)$. With these choices, the Langevin
equation~\eqref{Langevin} is mapped onto a new equation
\begin{equation}
\partial_{\tau} W_{\perp}(\tau)=-\Big(\frac{v_{th}}{\lambda_0}\tilde{\nu} \tilde{k}_m^2 -\omega_0\Big) W_{\perp}(\tau)+\sqrt{Nv^3_{th}\frac{c\tilde{\nu} \tilde{k}_m^2}{\lambda_0}  }~\xi(\tau),
\label {Lan-resc2}
\end{equation}
where $\lambda_0=L^2/(N\sigma)\equiv 1/(n\sigma)$ is proportional to
the mean free path and $\tilde{\nu}=\nu(t)/[\lambda_0 v_{th}(t)]$ and
$\tilde{k}_m=k_m \lambda_0$ are dimensionless rescaled viscosity and
wave number, respectively.

If $\tilde{\nu} \tilde{k}^2_m -\omega_0<0$, which is equivalent to
the condition of stability of shear modes in the HCS, the above
equation leads to a dimensionless autocorrelation function given
by
\begin{equation}
C_{\perp}(\tau)\equiv \frac{\langle W_{\perp}(0)W^*_{\perp}(\tau)\rangle} {\vt^2} =
\frac{cN}{2} \Big(1+\omega_0 \tau_0 \Big) e^{-\tau/\tau_0}
\label{qad-resc}
\end{equation}
with $\tau_0^{-1}\equiv \nt\kt_m^2\vt/\lambda_0-\omega_0$ the characteristic time of decay.\\
Based on Eq. (\ref{qad-resc}), we can obtain the dimensionless
kinematic viscosity $\nt$ and the ``FDR violation'' coefficient $c$ from a measure of $C_\perp(\tau)$:
\begin{eqnarray}
\nt&=&\frac{\lambda_0}{\vt\kt^2}\Big(\frac{1}{\tau_0}+\omega_0\Big) \label{nt}\\
c&=&\frac{2C_{\perp}(0)}{N\Big(\tau_0\omega_0+1\Big)} \label{c},
\end{eqnarray}
i.e. $\nt$ and $c$ are obtained measuring the amplitude $C_{\perp}(0)$ and the decay time $\tau_0$.

\section{Numerical results} \label{sec:numres}

The aim of this section is to measure $\nt$ and $c$ by means of Equations~\eqref{nt} and \eqref{c}
in DSMC simulations, comparing them to the results of MD. For this
purpose a DSMC algorithm must be carefully devised to guarantee the
validity of the HCS assumptions and to allow for the measure of
$W_\perp(\tau)$. This is done in the next subsection.

\subsection{Stationary representation for the inhomogeneous DSMC}
\label{statdsmc}

The DSMC algorithm used here is explained in details in the
Appendix. The essence of a DSMC algorithm, which has been originally
designed as a tool to numerically solve Boltzmann equations, is a
stochastic computation of collisions: pairs of particles and their
orientation vector $\hat{\sigma}$ are chosen randomly, with
probabilities dictated by the Boltzmann collision integral. In
homogeneous configurations, a pair of particles is chosen among all
$N$ particles of the system, disregarding spatial coordinates: the
large number $N$ guarantees Molecular Chaos. Here, to avoid shear mode
instabilities, we keep the total size of the system under the critical
size $L_c$, expecting to observe a homogeneous regime.  Nevertheless,
the quantity under scrutiny, $U_\perp(t)$, requires that collisions
are treated with a good spatial resolution, i.e. colliding particles
must be close to each other to give the correct contribution to the
variation of $U_\perp(t)$. For this reason, a inhomogeneous DSMC
algorithm is necessary. The system is partitioned in $m_c$
non-overlapping cells of size $l_c$: during the free-streaming step
each particle can move from a cell to a neighboring one, but during
the collision step the particle can collide with particles in the same
cell only. The resolution for the measure of $U_\perp(t)$ is improved
increasing $m_c$; at the same time, when $m_c$ increases, the average
number of particles in each cell $N_c=N/m_c$ decreases, threatening
the Molecular Chaos assumption. In principle the perfect resolution
would be achieved with $l_c \sim \sigma$, but this would result in a
number of particles per cell $N_c \sim n\sigma^2$ which, for
diluteness, is required to be much smaller than $1$. We will see in
the following (Fig.~\ref{fig2}) that taking $l_c \sim 15\sigma$ is
sufficient to restore a good resolution for measuring
$U_\perp(t)$. Such a choice, however, at the chosen packing fraction
$n\sigma^2=10^{-2}$ gives $N_c \sim 2.5$, which is so low that
the Molecular Chaos could be invalidated. A first check is to study the sensitivity 
of the rescaling procedure (\ref{resc}) (tested in \cite{BRM04} in the
homogeneous DSMC simulations and in~\cite{BRM98}) to the parameter
$m_c$ (or equivalently $N_c$).
  
\begin{figure}[htbp]
\begin{center}
\includegraphics[angle=0,width=9cm,clip=true]{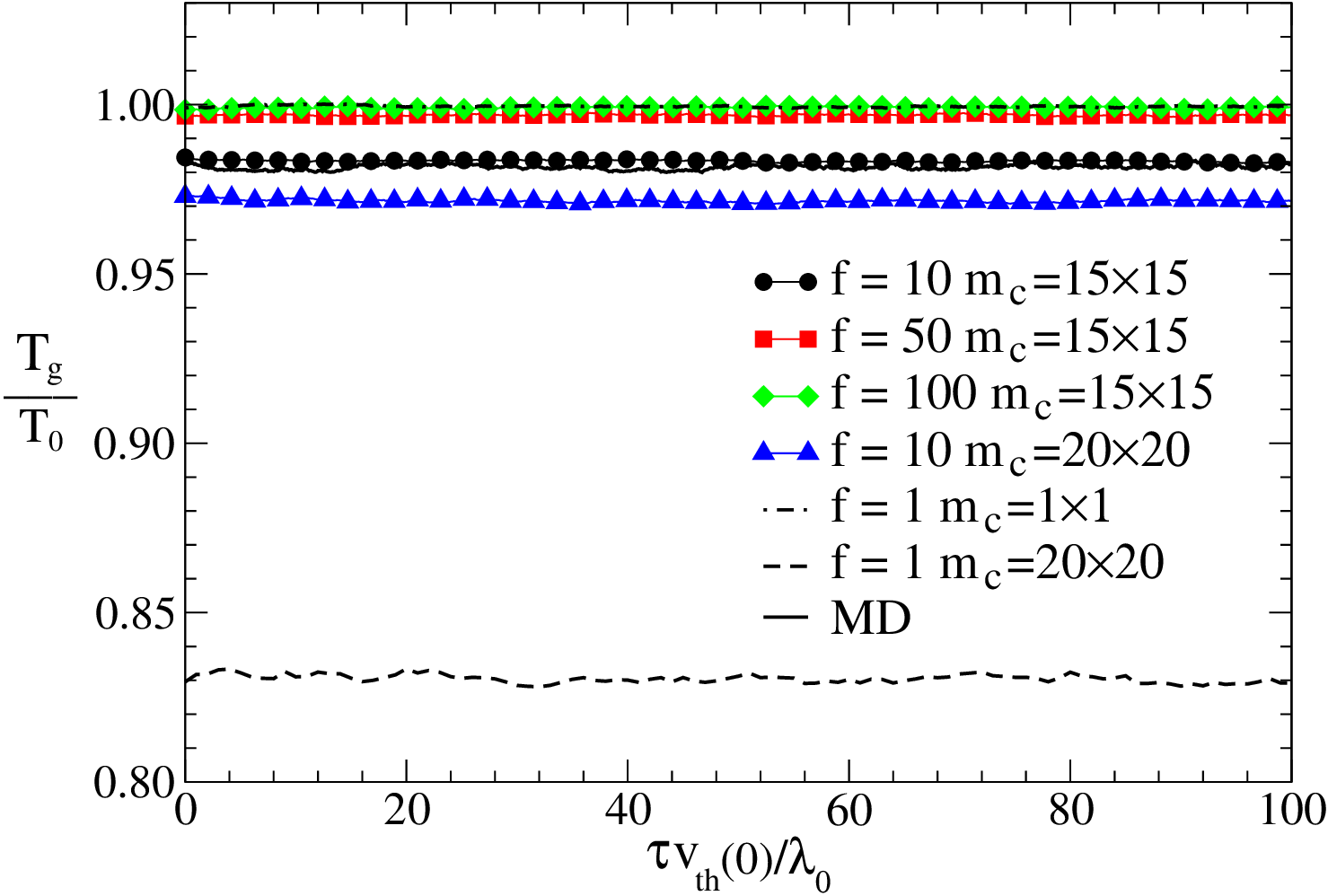}
\caption{The rescaled granular temperature $T_g$ as function of the
  rescaled time, in DSMC simulations with different $f$ factors and
  cell numbers $m_c$, compared with MD simulations, in the case of
  $N=1000$, $\alpha=0.8$ and $n\sigma^2=0.01$.  
\label{fig1}}
\end{center}
\end{figure}

\begin{figure}[htbp]
\begin{center}
\includegraphics[angle=0,width=8.5cm,clip=true]{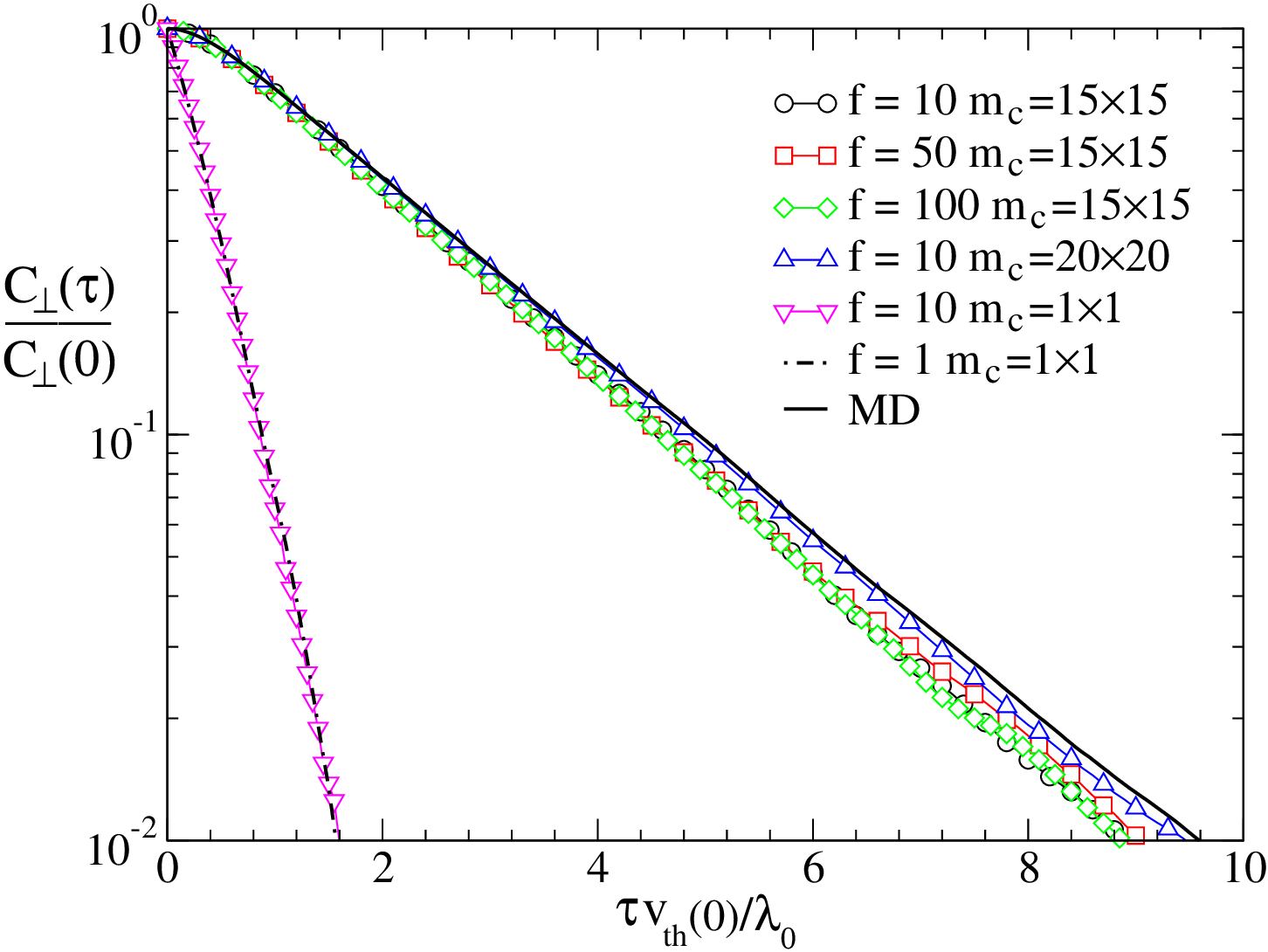}
\caption{The rescaled correlation function $C_{\perp}(\tau)$ (see Eq. \ref{qad-resc}) as function of the rescaled time, obtained from DSMC simulations with different $f$ factors and cell numbers $m_c$, and from MD simulations, in the case of $N=1000$, $\alpha=0.8$ and $n\sigma^2=0.01$.
\label{fig2}}
\end{center}
\end{figure}

Having this aim, we have characterized the entire system analyzing its
granular temperature $T_g$ as function of the rescaled time $\tau$ for
different choices of the cell number $m_c$ at $\alpha$ and $N$ fixed
(see Fig. \ref{fig1}), comparing it with MD results. The time-rescaling
is applied with a choice of $\omega_0$ such that, in the steady state
it is expected $T_g=T_0$. From the simulation data it results that a
steady state is always reached, i.e. the granular temperature is
constant in all cases. A homogeneous DSMC (i.e. $m_c=1$) gives an
agreement with the expected value which is better, of a few
percentages, with respect to MD. When $m_c$ is increased, however, the
DSMC gives very bad results. This is due to the small value of $N_c$,
which results in strong finite size effects, for instance fake
re-collisions which invalidate the Molecular Chaos hypothesis.

The way of solving this problem is to simulate a larger number of
particles $\tilde{N}=fN$ with $f>1$, keeping the collision statistics
per particle as that of a $N$-particles system (see the Appendix for a
precise description). This is equivalent to simulate $f$ copies of the
original $N$-particles system and let particles of different copies
swap at each time step. The effect of $f>1$ is immediately seen in
Fig.~\ref{fig1}: the steady state granular temperature perfectly
agrees with that measured in MD, as soon as $f\sim 10$, even if
$m_c=225$. 
Much larger values of $f$ appear to slightly
improve the value of $T_g$. On the other side a too much large
value of $f$ can be very expensive in terms of cpu-time.

A crucial quantity, here, is the number of virtual
particles per cell $\tilde{N}_c=fN_c=fN/m_c$. If $m_c$ is increased
(for instance in Fig.~\ref{fig1} it is changed from $225$ to $400$),
$f$ should be increased. Apparently it is sufficient to have
$\tilde{N}_c \sim 25$ to have a good agreement, for $T_g$, between DSMC and
MD. 

As discussed above, the necessity of a large number of cells comes
from the sensitivity of $U_\perp(t)$ to the spatial localization of
collisions. This is shown in Fig.~\ref{fig2}, where the decay of
$C_\perp(\tau)$ is displayed: the homogeneous DSMC ($m_c=1$) shows an
exponential decay with a $\tau_0$ completely different from the one
observed in the MD simulations. This behavior  is insensitive
to changes of $f$ from 1 to 10. In order to obtain the correct decay, it
is necessary to use the inhomogeneous DSMC approach with $m_c \gg 1$. A
partition of $m_c=20\times 20$ for $N=1000$, in fact, reveals an exponential decay very
close to the MD results: also in this case, a change of $f$ from
10 to 100 is unimportant.

\begin{figure}[htbp]
\begin{center}
\includegraphics[angle=0,width=8.5cm,clip=true]{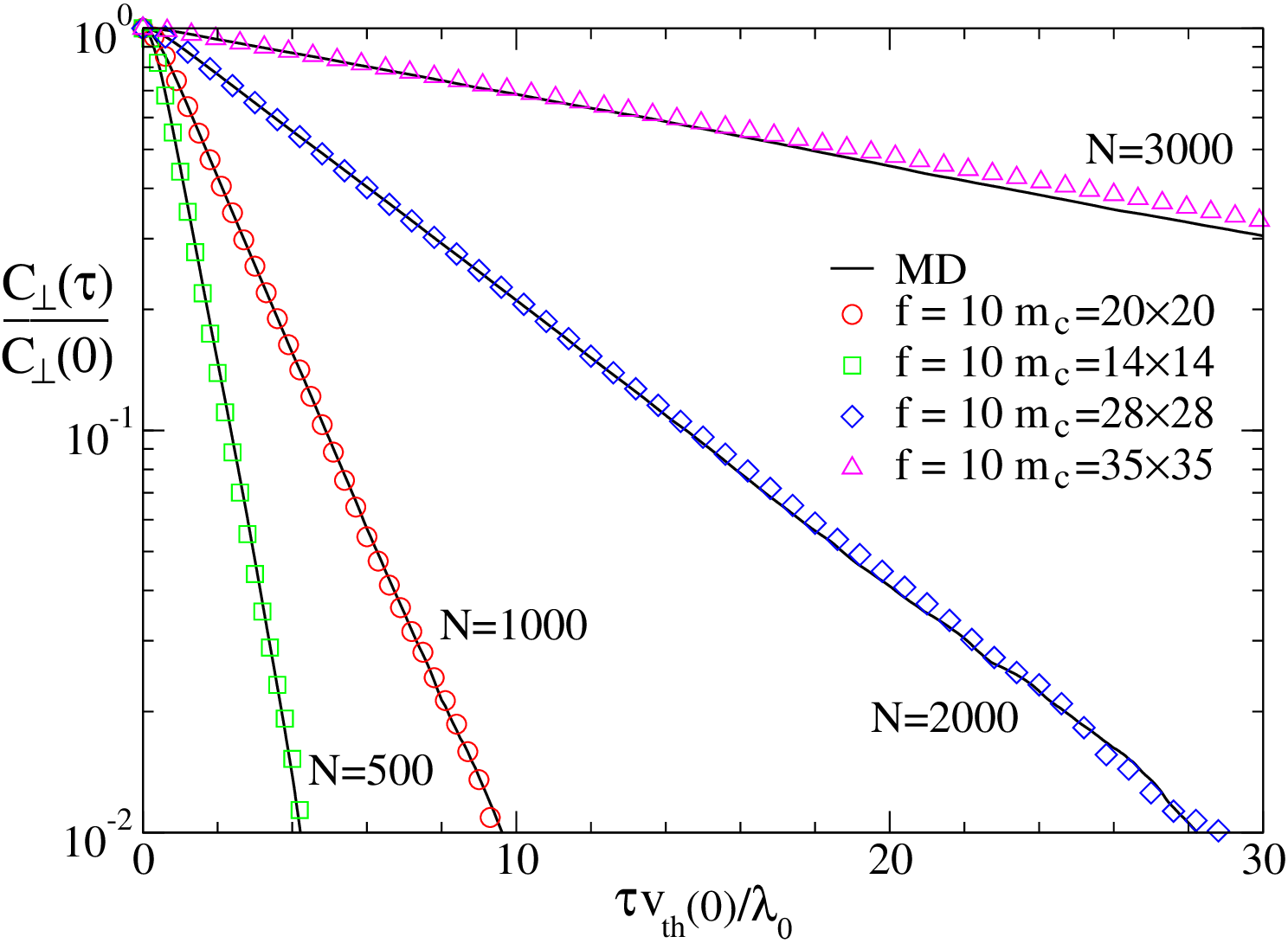}
\caption{The rescaled correlation function $C_{\perp}(\tau)$ as
  function of the rescaled time, from DSMC simulations with
  different total particle numbers $N$ and same average number of
  particles per cell $\tilde{N}_c$. Data from MD simulations are shown
  for comparison. The parameters are the same as in
  Fig. \ref{fig1}.  \label{fig3}}
\end{center}
\end{figure}

When $N$ is increased at fixed density (and mean free path), the size
of cells must be kept constant, which implies $m_c \propto N$. If $f$
is also kept constant, $\tilde{N}_c$ is automatically preserved and we
expect to have a good comparison with MD simulations in terms of
Molecular Chaos, as well as decay of $C_\perp(\tau)$. This situation
is fairly verified in Fig.~\ref{fig3}, where the decay of
$C_\perp(\tau)$ is shown to be always close to the MD results.

This approach is valid in general and, with proper values of $f$ and
$m_c$, we can compare, for different values of $\alpha$ and $N$, the
properties of the fluctuations of $W_\perp$ between DSMC and MD
simulations.  Guided by this preliminary analysis, we have set the
value of $f=10$ and $\tilde{N}_c=25$ for all the following discussion.

\subsection{Analysis of the fluctuations}

In the previous section we made the hypothesis that the fluctuations of
the slowest mode of the transverse velocity field $U_\perp(t)$ follow
the Langevin equation (\ref{Langevin}), or (\ref{Lan-resc2}) in the
rescaled representation: this implies that its correlation function
follows the exponential decay in Eq. (\ref{qad-resc}), with
characteristic time $\tau_0$.

\begin{figure}[htbp]
\begin{center}
\includegraphics[angle=0,width=9cm,clip=true]{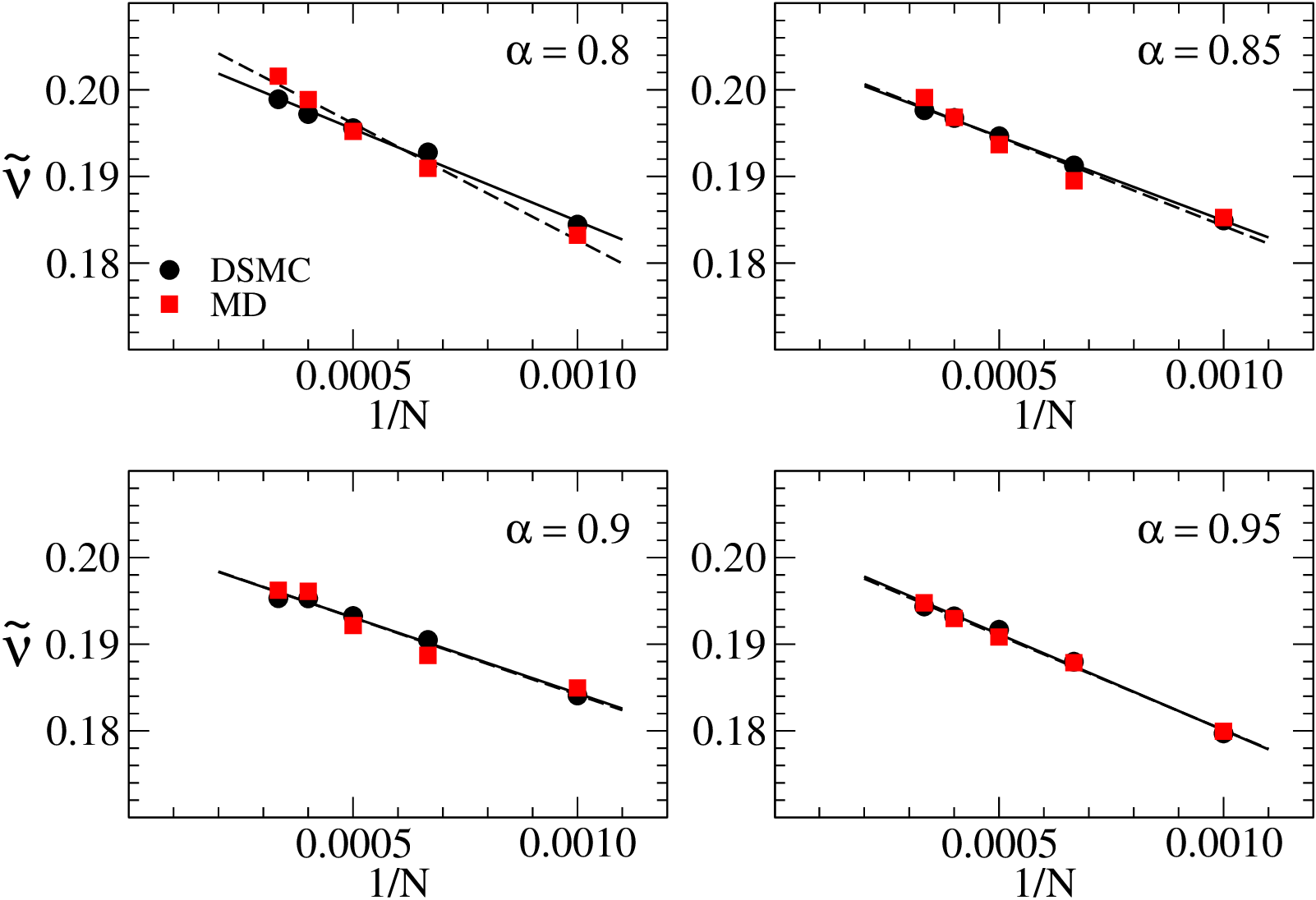}
\caption{The dimensionless kinematic viscosity $\nt$, obtained from
  Eq. \eqref{nt} and from the exponential decay of the correlation function
  in Eq. (\ref{qad-resc}), as function of the inverse of the number of
  particles $N$ for different values of $\alpha$. The circles and
  squares correspond to data obtained from DSMC an MD simulations,
  respectively. The solid and dashed lines are the best linear
  fit. The density is the same as in Fig. \ref{fig1}.\label{fig4}}
\end{center}
\end{figure}

In order to apply the correct thermostat
$\dot{\tilde{\mathbf{v}}}=\omega_0 \tilde{\mathbf{v}}$, and to obtain the
kinematic viscosity $\nt$ from Eq.(\ref{qad-resc}), it is necessary to
know the value of $\omega_0=\zeta_H(0)/2$. This is computed, in the homogeneous
case, giving the theoretical expressions~\cite{BRM04}
\begin{equation}
\zeta_H(0)=\frac{\vt(0)}{\lambda_0}\sqrt{\frac{\pi}{2}}\Big[1+\frac{3}{16}
a_2(\alpha)\Big]
\label{zeta}
\end{equation}
where the coefficient $a_2(\alpha)$ is 
\begin{equation}
a_2(\alpha)=\frac{16(1-\alpha)(1-2\alpha^2)}{57-25\alpha+30\alpha^2-30\alpha^3}.
\end{equation}
We have verified that the prediction in Eq. (\ref{zeta}) is good also
for the inhomogeneous DSMC and for MD.

Eq.~\eqref{qad-resc} is expected to be valid in the large $N$ limit:
for this reason we have performed simulations for different values of
$N$ (1000, 1500, 2000, 2500 and 3000), fixing the particle density $n$ of
the system, and measuring the time decay $\tau_0$. The data obtained
using~\eqref{nt} and~\eqref{zeta} in DSMC and MD, are shown in
Fig. \ref{fig4}. The dimensionless viscosity $\nt$ appears to be a
linear function of $N^{-1}$, therefore a best linear fit allows to
extrapolate the asymptotic value for $N\to\infty$. The agreement
between MD and DSMC data is good in all cases, even at not too high
values of $N$.

\begin{figure}[htbp]
\begin{center}
\includegraphics[angle=0,width=9cm,clip=true]{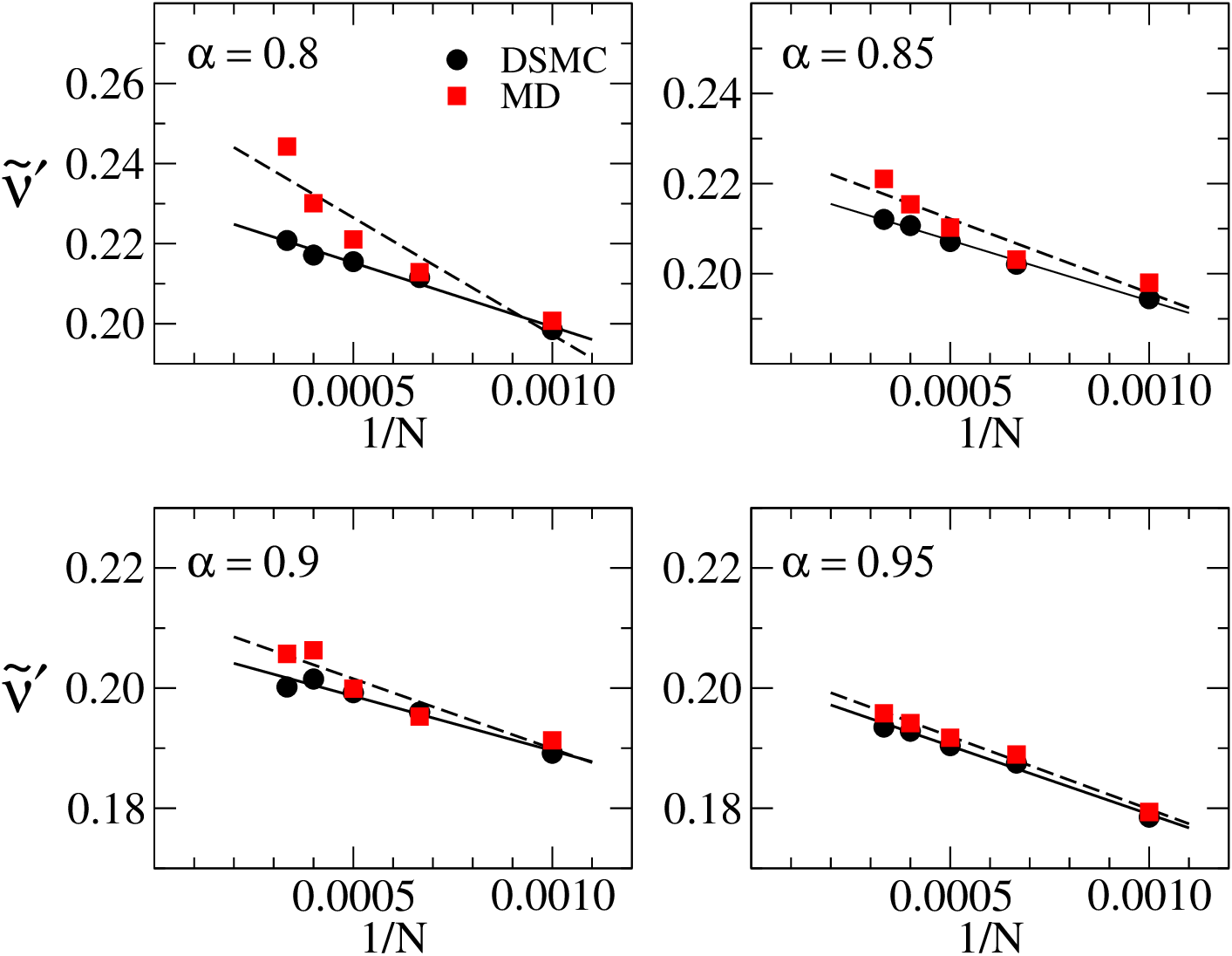}
\caption{The dimensionless viscosity $\nt'=c\,\nt$ obtained from
  Eq. (\ref{c}) and (\ref{nt}) (see text), as function of the inverse
  of the number of particles $N$ for the following values of
  $\alpha$: 0.8, 0.85, 0.9 and 0.95. The circles and squares
  correspond to data obtained from DSMC an MD simulations,
  respectively. The solid and dashed lines are the best linear
  fit. The density is the same as in Fig. \ref{fig1}.  \label{fig5}}
\end{center}
\end{figure}

Our second question is if the inhomogeneous DSMC can reproduce the MD
results for the noise term in the Langevin equation (\ref{Langevin})
or (\ref{Lan-resc2}), in particular the quantity $\nt'\equiv c\nt$
that is the viscosity term appearing in the noise. Analogously to
Fig. \ref{fig4}, we have analyzed its dependence versus $N^{-1}$. Also
in this case we have found a linear law, shown in Fig. \ref{fig5}: a
small discrepancy is observed, for this quantity, between DSMC and MD
data, when $N$ is increased.  The disagreement results more evident if
the collisions are more inelastic. Our interpretation of this small
difference is that the DSMC dynamics at small time-scales is
  slightly different from MD: indeed, this disagreement is less
evident at large time-scales, for instance in the measure of $\nt$.
If we consider also that the DSMC and MD results about the
  correlation of the energy fluctuations are very
  similar~\cite{VPBWT06}, we tend to conclude that this small
  discrepancy could be due to the use of small cells (large $m_c$ or
  equivalently small $\tilde{N}_c$), which is necessary in the
  measures of $C_{\perp}(0)$. A more general discussion of the
  cause of this discrepancy is given at the end of this section.

\begin{figure}[htbp]
\begin{center}
\includegraphics[angle=0,width=9cm,clip=true]{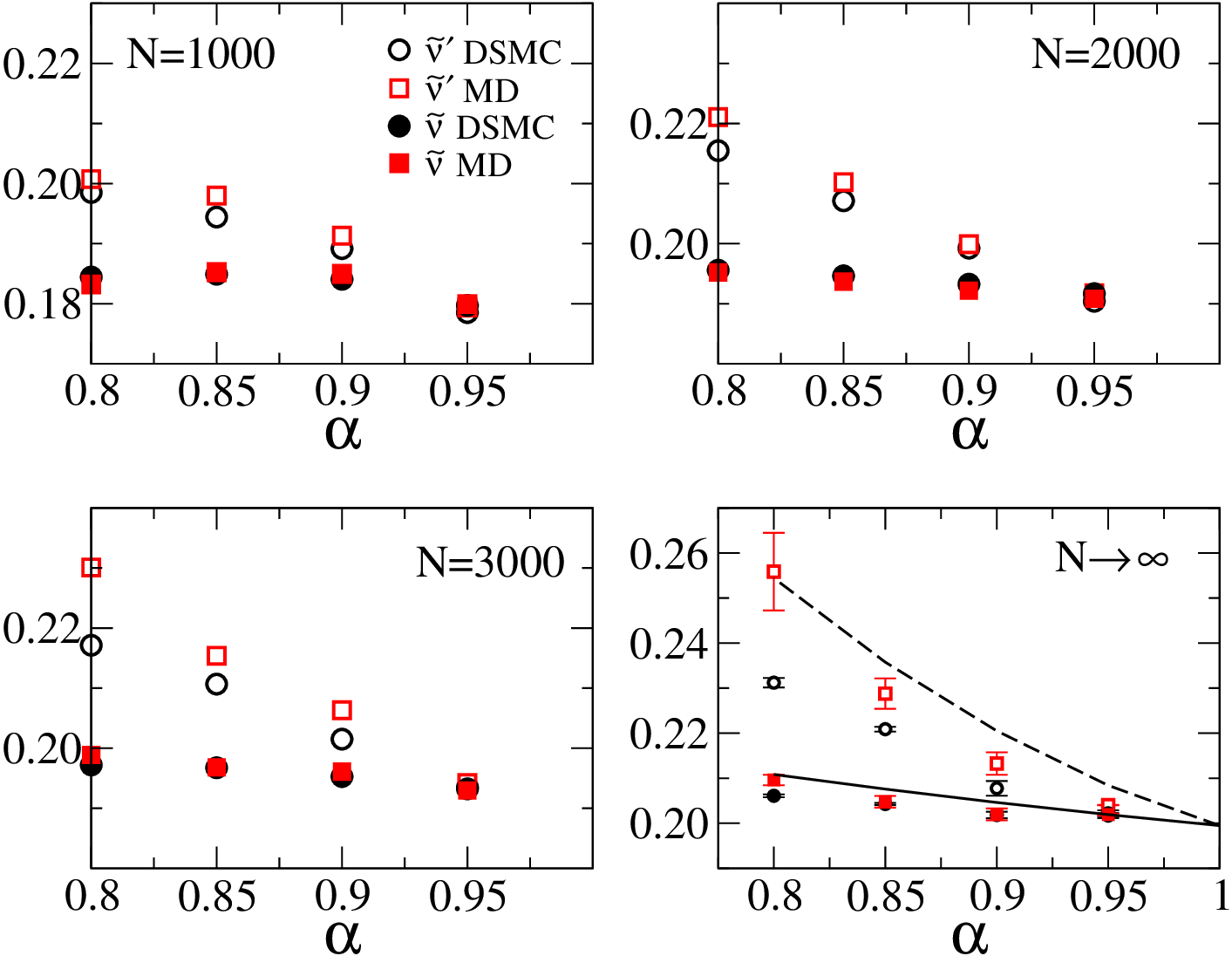}
\caption{The dimensionless viscosity $\nt'=c\,\nt$ (empty symbols) and
  the dimensionless shear viscosity $\nt$ (full symbols) as function
  of the coefficient of restitution $\alpha$ for different values of
  $N$: 1000, 2000 and 3000. The circles and squares correspond to DSMC
  and MD data, respectively.  The data for $N\to\infty$ are
  extrapolated, with the respective errors, from the linear fits of
  Fig. \ref{fig4} and Fig. \ref{fig5}. The solid and dashed lines are
  the theoretical predictions \cite{BMG09}.\label{fig6}}
\end{center}
\end{figure}

Notwithstanding this slight discrepancy, the value of $c=\nt'/\nt$
results significantly larger than $1$ also in DSMC.  The inhomogeneous
DSMC is then able to reproduce the breakdown of the FDR as already
found with the MD approach \cite{BGM08}. This is evident from
Fig.~\ref{fig6}, where the $\nt$ and $\nt'$ are plotted as function of
$\alpha$ for different values or the particle number and for
$N\to\infty$.  The extrapolated results for $N\to\infty$, for $\nt$
are in good agreement with the MD data and with the theoretical
analysis of~\cite{BMG09}, while those for $\nt'$ still underestimate
them, as shown in the fourth panel of Fig. \ref{fig6}. 

\begin{figure}[htbp]
\begin{center}
\includegraphics[angle=0,width=9cm,clip=true]{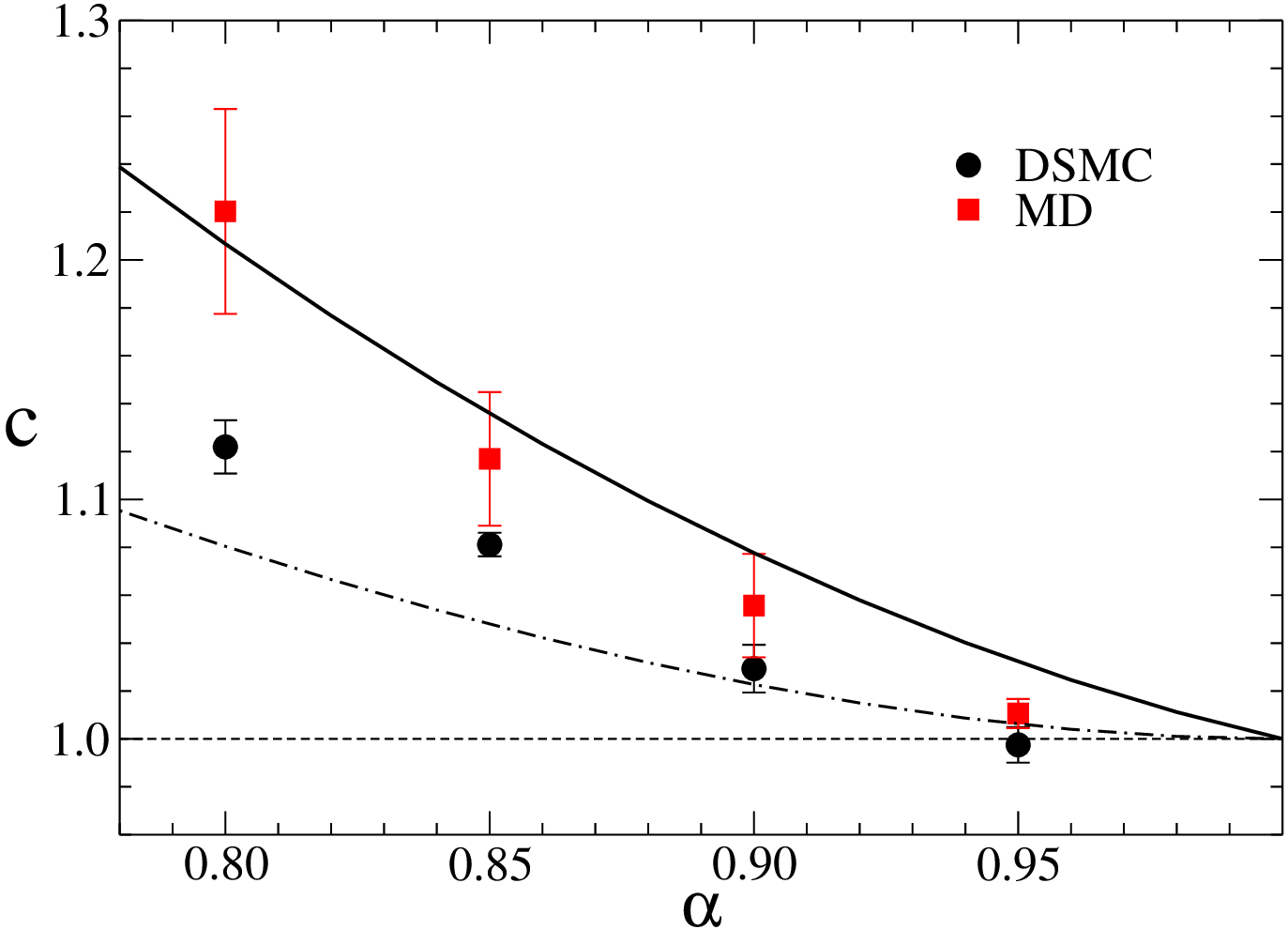}
\caption{The ratio $c=\nt'/\nt$ as function of the coefficient of
  restitution $\alpha$ for $N\to\infty$ (see fourth panel of
  Fig. \ref{fig6}).  The circles and squares correspond to DSMC and MD
  data, respectively. The solid is the theoretical prediction
  \cite{BMG09}. The value of reference $c=1$ (dashed line) is obtained
  if FDR are satisfied.  The dot-dashed curve is the
  result of the theoretical analysis of~\cite{BMG09} with
  the additional assumption of white noise.
\label{fig7}}
\end{center}
\end{figure}

The ratio $c$ is shown in Fig.~\ref{fig7}: it clearly displays strong
deviations from $1$ (violations of FDR), which reach $12\%$ at
$\alpha=0.8$ also in DSMC.  The dot-dashed curve represents the
  result for $c$ from the theoretical analysis of~\cite{BMG09} with
  the additional assumption of white noise. We note a close comparison
  with the data from DSMC, suggesting that the effect of this protocol
  is an effective time-decorrelation of noise. Such a conclusion
  should be taken with care, in view of other simulations, performed
  with different choices of $f$ and $m_c$: these results are not shown
  here, since have a quite high statistical uncertainty due to the
  large computational time required (increasing $f$ and $m_c$ results
  in an increase of fictive particles $\tilde{N}$). They suggest that,
  when $f$ and $m_c$ are increased keeping $\tilde{N}_c=fN_c$
  constant, the value of $c$ stays substantially unchanged. On the
  other side if $\tilde{N}_c$ is increased (e.g. by increasing $f$ at
  constant $m_c$) the value of $c$ slightly increases and seems to
  slowly tend toward the corresponding MD value, which is also
  compatible with the theoretical analysis with a {\em colored}
  noise. The presence of time-correlation in the noise, even in DSMC, is in
  agreement with the results for $C_\perp(\tau)$ shown in
  Fig.~\ref{fig2} and~\ref{fig3}, which are not straight exponentials,
  but some bending can be observed at small $\tau$.

Apart from these small discrepancies, the main fact is that the
inhomogeneous DSMC is able to reproduce the violation of FDR, which was not
obvious. In fact, it is reasonable and well verified that average
values (e.g. transport coefficients) obtained in DSMC agree with
dilute MD simulations; anyway it is less trivial to observe good
agreement for fluctuations, which - in non-equilibrium situations -
can be much more sensitive to the detailed mechanisms of the dynamics.

\section{Conclusions} \label{sec:conc}

In this work we have obtained two main results. First, we have
individuated which are the basic ingredients - in a Direct Simulation
Monte Carlo algorithm - to correctly measure the time-decay of the
autocorrelation function of fluctuations of a hydrodynamic field,
keeping valid the assumption of Molecular Chaos: the volume must be
divided in small cells for the purpose of collision computation, even
if the system is assumed to be spatially homogeneous; the good
resolution to appreciate the ``real'' time-decay of $C_\perp(\tau)$ is
achieved only when $N_c \sim 2$ which is too small to avoid fake
recollisions: for this reason, a number (order $\sim 10$) of
``virtual'' copies of the system is necessary to restore Molecular
Chaos. This procedure is similar to that adopted in~\cite{BR04}. 
 The study of the transverse velocity field could also be done by
  dividing the system in slides, as in~\cite{BRM07,GMLMC87}. Such a
  procedure can be used to analyze $U_{\perp}(\tau)$, while we have no
  knowledge of its adaptation to other space-dependent fields. Note
that - when DSMC is used to study the homogeneous cooling state,
i.e. keeping the system size below the critical size for instability,
$L<L_c$ - one usually does not divide the system in cells, and still
obtains excellent agreement for one-point observables. Previous study
of cooling granular gases with inhomogeneous DSMC~\cite{BRM98,BR04}
where in fact proposed to analyze the {\em departure} from the HCS.

Our second result concerns a quantitative agreement with Molecular
Dynamics results, which is very good for the exponential tail of the
autocorrelation of fluctuations. The amplitude of fluctuations, on the
other side, shows small discrepancies with respect to MD results,
which become more evident as $\alpha$ is reduced and $N$ increases. In
the extrapolated $N\to \infty$ limit, these discrepancies can reach
(for $\alpha=0.8$) an underestimation of order $10\%$: nevertheless it
is still possible to appreciate a $\sim 12\%$ of deviation from the
validity of FDR, which is one of the most peculiar property of
granular hydrodynamic fluctuations in the HCS. In DSMC simulations,
deviations from a perfect exponential decay of $C_\perp(\tau)$ are
also visible, at small times, but these are only qualitatively similar
to those observed and predicted in~\cite{BMG09}.

Hydrodynamic noise is always important in granular systems, which - in
terms of number of elementary constituents - are much smaller than
molecular fluids ($10^4$ instead of $10^{20}$ particles). Future
research is needed to understand the properties of granular
hydrodynamic noise in steady state models: even if we have shown that
these properties do not depend on the detailed collisional mechanism,
it seems that they depend on the choice of energy driving protocol~\cite{MST}.

\section*{Acknowledgments}

The authors are indebted with A. Sarracino for useful discussions
and a careful reading of the manuscript. The work of the authors is
supported by the ``Granular-Chaos'' project, funded by the Italian
MIUR under the FIRB-IDEAS grant number RBID08Z9JE.

\appendix
\section*{Appendix}
 \label{app}

In this Appendix we give a detailed description of the DSMC algorithm
used in this work. The algorithm involves the
simulation of $\tilde{N}=fN$ fictive particles with $f>1$. This is a
trick often used in DSMC: for molecular gases - where $N\sim 10^{20}$ - it
is customary to use $f \ll 1$ in order to have a manageable number of
fictive particles. In granular gases, where the number of real particles is
small, one uses $f>1$ in order to have enough particles in a cell to
guarantee Molecular Chaos. This is our case.

As initial condition we have chosen a uniformly random spatial
configuration of the particles and a Gaussian distribution of their
velocities.  After initialization, the dynamics consists of a cycle
with the following two steps:
\begin{itemize}
\item Streaming step (evolution of positions and velocities ignoring possible collisions)
 \item Collision step
\end{itemize}

The {\bf streaming step} consists in computing, for every particle $i$, the
time-discretized version (with a fixed $\Delta t$, small enough) of
the evolution equations

\begin{align}
\dot{\bf r}_i&=\tilde{\bf v}_i\\
\dot{\tilde{\mathbf{v}}}_i&=\omega_0 \tilde{\mathbf{v}}_i
\end{align}

For the purpose of the {\bf collision step}, the system is partitioned
in $m_c$ equal non-overlapping cells of size $l_c$ (an optimal value
$l_c \sim 15\sigma$ has been individuated, as discussed in
Section~\ref{statdsmc}). The average number of fictive particles in
the cell $i$ is then $N_c=\tilde{N}/m_c$. If $m_c$ is too large, the
number of real particles in a cell $N/m_c$, results too small and
invalidates the Molecular Chaos hypothesis, due to fake
recollisions. This is avoided by taking $f>1$ large enough (a value of
$\sim 10$ seems optimal).

In each cell $k$ the local average collision frequency - per particle
- $\omega_k=\sigma n_k\sqrt{4\pi T_k/m}$ is calculated, based on
  the {\em real} density $n_k=\tilde{N}_k/(f l^2_c)$ where
  $\tilde{N}_k$ is the number of fictive particles in the cell and an
estimate of the local temperature $T_k$ (i.e. the variance of the
local particle velocity distribution). Then a number of collisions
$\tilde{N}_k \omega_k \Delta t$ is performed: this is obtained
choosing, at random, pairs of particles $i$ and $j$, and unit vector
(uniformly distributed angle) $\hat{\sigma}$, accepting them with a
probability proportional to $(\bf v_i-\bf v_j)\cdot \hat{\sigma}$. The
accepted pair is updated with equations~\eqref{rules}.

The algorithm described here is the classical DSMC algorithm as
proposed by Bird~\cite{B94} (variant exist which include fluctuations
of the number of collisions with the introduction of an internal clock
for each cell). It is interesting, in this particular case, to realize
that the number $f$, when integer and larger than $1$, can be thought
as a number of ``copies'' of the system. In fact, if $\Delta t$ is
small enough to guarantee that $N_k \omega_k \Delta t \ll N_k$ for any
cell $k$ (which is always true in our simulations), then one can
always {\em virtually} separate the $\tilde{N}_k$ fictive particles of
the cell into $f$ groups of - averagely - $N_k$ particles, such that
collisions occur only between particles of the same group. Given that
colliding pairs are chosen at random among the $\tilde{N}_k$
particles, at each step this virtual separation into $f$ groups is
done in a new way, which is equivalent to say that - at each
step - some of the particles move from one group to the other,
remaining of course in the same cell.
 
This is just an alternative representation of the algorithm, which
underlines the randomizing mechanism - intrinsic of DSMC and not
present in MD.

\bibliography{fluct.bib}

\end{document}